\def\ref{\par\noindent\hang}

\def\spose#1{\hbox to 0pt{#1\hss}}
\def\approxlt{\mathrel{\spose{\lower 3pt\hbox{$\sim$}}
        \raise 2.0pt\hbox{$<$}}}
\def\approxgt{\mathrel{\spose{\lower 3pt\hbox{$\sim$}}
        \raise 2.0pt\hbox{$>$}}}

\def\multleft#1{\hbox to size{\vbox {\halign {\lft{##}\cr #1}}\hfill}\par}
\def\multright#1{\hbox to size{\vbox {\halign {\rt{##}\cr #1}}\hfill}\par}

\def\$<${\thinspace}
\def\s{\hbox{\phantom{5}}}      %one space
\def\boxit#1{\vbox{\hrule\hbox{\vrule\kern3pt\vbox{\kern3pt
          #1 \kern3pt}\kern3pt\vrule}\hrule}}

\def\cm{{\rm\thinspace cm}}

\def\erg{{\rm\thinspace erg}}

\def\g{{\rm\thinspace g}}

\def\km{{\rm\thinspace km}}
\def\kpc{{\rm\thinspace kpc}}

\def\pc{{\rm\thinspace pc}}

\def\s{{\rm\thinspace s}}

\def\ergps{\hbox{$\erg\s^{-1}\,$}}
\def\gpcm{\hbox{$\g\cm^{-3}\,$}}

\def\kmps{\hbox{$\km\s^{-1}\,$}}

\def\rc{\hbox{$r_{\rm c}$}}
\def\rs{\hbox{$r_{\rm s}$}}
\def\Vs{\hbox{$V_{\rm s}$}}
\def\Vc{\hbox{$V_{\rm c}$}}
\def\gc{\hbox{$\gamma_{\rm c}$}}
\def\gs{\hbox{$\gamma_{\rm s}$}}

\documentclass{article}

\usepackage{psfig}
\usepackage{aaspp4}
\usepackage{times}
\usepackage{amssym}

\makeatletter
\@ifundefined{chapter}{\def\thebibliography#1{
\section*{References}
  \list
  {\relax}{\setlength{\labelsep}{0em}
        \setlength{\itemindent}{-\bibhang}
        \setlength{\itemsep}{\parskip}
        \setlength{\parsep}{0pt}
        \setlength{\leftmargin}{\bibhang}}
    \def\newblock{\hskip .11em plus .33em minus .07em}
    \sloppy\clubpenalty4000\widowpenalty4000
    \sfcode`\.=1000\relax}}%
{\def\thebibliography#1{
  \list
  {\relax}{\setlength{\labelsep}{0em}
        \setlength{\itemindent}{-\bibhang}
        \setlength{\itemsep}{\parskip}
        \setlength{\parsep}{0pt}
        \setlength{\leftmargin}{\bibhang}}
    \def\newblock{\hskip .11em plus .33em minus .07em}
    \sloppy\clubpenalty4000\widowpenalty4000
    \sfcode`\.=1000\relax}}
 
\newlength{\bibhang}
\let\@internalcite\cite
\def\cite{\@ifstar{\citey}{\citefull}}
\def\citefull{\def\astroncite##1##2{##1\ ##2}\@internalcite}
\def\citey{\def\astroncite##1##2{##1\ (##2)}\@internalcite}
\def\citeyear{\def\astroncite##1##2{##2}\@internalcite}
\def\citename{\def\astroncite##1##2{##1}\@internalcite}
\def\@citex[#1]#2{\if@filesw\immediate\write\@auxout{\string\citation{#2}}\fi
  \def\@citea{}\@cite{\@for\@citeb:=#2\do
    {\@citea\def\@citea{; }\@ifundefined
       {b@\@citeb}{{\bf ??}\@warning
       {Citation `\@citeb' on page \thepage \space undefined}}%
{\csname b@\@citeb\endcsname}}}{#1}}
 
\def\@cite#1#2{#1\if@tempswa #2\fi} 
\def\@biblabel#1{}
 
\def\astroncite#1#2{#1\ #2}
\makeatother

\begin{document}

\title{Intermittent radio galaxies and source statistics}

\author{Christopher S. Reynolds\altaffilmark{1} and Mitchell C. Begelman\altaffilmark{1,2}}
 
\altaffiltext{1}{JILA, University of Colorado, Campus Box 440, Boulder, CO 80309-0440\\
\{chris,mitch\}@rocinante.colorado.edu}
 
\altaffiltext{2}{Department of Astrophysical and Planetary Sciences, \\ 
University of Colorado, Boulder, CO 80309-0391.}

\begin{abstract}
We suggest that extragalactic radio sources are intermittent on timescales
of $\sim 10^4$--$10^5$\,yr.  Using a simple spherical model of a
cocoon/shock system, it is found that inactive sources fade rapidly in
radio luminosity but the shock in the ambient medium continues to expand
supersonically, thereby keeping the whole source structure intact during
the inactive phases.  The fading of inactive sources, and the effect of the
intermittency on the expansion velocity, can readily explain the observed
over-abundance of small radio sources.  In particular, the plateau in the
observed distribution of sizes found by O'Dea \& Baum (1997) can be
interpreted as being due to intermittency.  The model predicts that very
young sources will be particularly radio luminous, once the effects of
absorption have been accounted for.  Furthermore, it predicts the existence
of a significant number of faint `coasting' sources.  These might be
detectable in deep, low-frequency radio maps, or via the X-ray and optical
emission line properties of the shock front.
\end{abstract}


\section{Introduction}

Recent radio surveys have identified numerous objects which are
morphologically similar to FR-II radio galaxies but are appreciably
smaller.  Those sources that are less than 500\,pc in extent are termed
Compact Symmetric Objects (CSOs; Wilkinson et al. 1994).  This class also
contains many Gigahertz Peaked Sources (GPSs), sources whose radio spectrum
is seen to peak at GHz frequencies (O'Dea, Baum \& Stanghellini 1991).  The
spectral form of GPSs is thought to be due to either free-free absorption
by an inhomogeneous foreground screen, or synchrotron self-absorption in
the source itself.  Slightly larger sources, those in the range
0.5--15\,kpc, have been termed Medium Symmetric Objects (MSOs) by Fanti et
al. (1995).  These various classes of small sources are found to constitute
10--30 per cent of all sources in a flux limited sample.

It is tempting to consider an evolutionary picture in which CSOs evolve
into full size FR-II radio galaxies, passing through the MSO stage.  Since
we would expect the sources to remain small for a relatively short period
of time, there must be strong luminosity evolution for us to see so many
small sources (Begelman 1996; Readhead et al. 1996).  Begelman (1996;
hereafter B96) showed how this luminosity evolution could be understood in
terms of a declining source pressure as the source evolves.

O'Dea \& Baum (1997; hereafter OB97) have recently studied a combined
sample of objects including CSOs, MSOs and classical FR-II radio galaxies.
In particular, they examine the distribution of (projected) linear sizes.
Using the B96 evolution model, and assuming physically realistic
interstellar medium (ISM) density profiles, they find a clear overabundance
of CSOs and MSOs as compared with the classical FR-II radio galaxies.  This
is seen as a `plateau' in the size distribution of their sample of sources
between $\sim 100\pc$ and $\sim 10\kpc$.  They suggest that either (1) a
large fraction of the small sources are transient or frustrated and never
evolve into large sources, or (2) the luminosity evolution is much stronger
than that predicted by B96, possibly due to a decline in the efficiency of
conversion of jet kinetic energy into radio power.

In this {\it letter} we suggest that radio sources are intermittent, and
that the source statistics examined by OB97 can be understood in the
context of a simple evolutionary picture if this intermittency is taken
into account.  In Section 2, we develop a simple model of radio source
evolution including intermittency.  Theoretical source statistics are
calculated in Section 3.  Section 4 discusses predictions of this scenario.
Our conclusions are summarized in Section 5.

\section{The coupled cocoon/shocked-shell model}

\subsection{The basic model}

Adopting the standard evolutionary picture (Scheuer 1974; Begelman \&
Cioffi 1989), we assume that the radio jets are enveloped in, and feed, a
cocoon of relativistic material which is overpressured with respect to the
ambient medium (which may be the ISM of the host galaxy, or the
intra-cluster medium (ICM) of the host cluster).  This overpressure drives
a strong shock into the ambient medium and forms a shell of shocked ISM/ICM
surrounding the relativistic cocoon.  The expansion velocity of this shell
is determined by the ram pressure of the ambient material entering the
shock.  The cocoon material and the shocked ISM/ICM shell are separated by
a contact discontinuity.

While it is clear that the large scale structures of radio sources show
elongation along the jet axis, one rarely observes large axial ratios.  For
example, low-frequency (327~MHz) radio maps of the powerful FR-II source
Cygnus~A (Carilli, Perley \& Harris 1994) reveal a radio emitting cocoon
with an axial ratio of $\sim 3$, even though its jets are observed to be
collimated to within a few degrees.  A similar situation is found for the
smaller CSOs and MSOs.  Thus, for the purposes of our simple model, we
shall assume that the cocoon and bow-shock are spherical.  We will denote
the radius of the cocoon as $\rc$, and the radius of the bow-shock
as $\rs$ ($\rs>\rc$).   We also denote by $\Vc$ and $\Vs$ the volume of the
cocoon and shocked shell, respectively.

We make several further assumptions.  Firstly, we assume that at any given
instant the pressure within the shock (of both the cocoon and the shocked
ISM/ICM shell) is spatially uniform with value $p(t)$.  See Kaiser \&
Alexander (1997) for an explicit justification of this assumption.
Secondly, we suppose that only a small fraction of the total kinetic
luminosity of the source, $L_{\rm j}(t)$, is radiated.  The rest of this energy is
assumed to be fed into the cocoon and thus drive the expansion of the
cocoon/shocked-shell system.  Thirdly, the ambient (undisturbed) medium is
assumed to have a density distribution of the form
$\rho(r)=\rho_0(r/a)^{-\alpha}$, where $r$ is the distance from the center
of the radio source.

Given these assumptions, the conservation of energy can be applied to the
cocoon and shocked shell to give,
\begin{equation}
(\gc-1)^{-1}(\Vc\dot{p}+\gc p\dot{\Vc})=L_{\rm j}(t),
\end{equation}
and
\begin{equation}
(\gs-1)^{-1}(\Vs\dot{p}+\gs p\dot{\Vs})={1\over 2}4\pi\rs^2 \rho(\rs) \dot{\rs}^3,
\end{equation}
where $\gc$ and $\gs$ are the ratio of specific heat capacities in the
cocoon and shocked-shell, respectively, and the dot denotes differentiation
with respect to time.  We close the system of equations with the
ram-pressure condition, $p=\rho(\rs)\dot{\rs}^2$.  This condition will be
valid provided that the expansion of the shocked shell remains highly
supersonic with respect to the ambient medium.  If the ambient medium is
identified with the hot component of the ISM/ICM, its sounds speed will be
$c_{\rm s}\sim 1000\kmps$.  Thus, the expansion of the source will be
highly supersonic provided $\dot{\rs}\approxgt {\rm few}\times 1000\kmps$.
Once the expansion of the source ceases to be supersonic, the
cocoon/shocked-shell structure will disrupt and dissipate.

In order to relate this model to radio observations, we need a prescription
relating the radio luminosity, $Q$, to the physical parameters of the
model.  To do this, we assume that the radio emission is dominated by the
synchrotron radiation of the relativistic electrons in the cocoon.  If we
further suppose that the magnetic field is in equipartition with the
relativistic electrons, standard minimum pressure arguments give $Q\propto
p^{7/4}\Vc$.

\subsection{Evolution of a periodically intermittent source}

\begin{figure*}[t]
\hbox{
\psfig{figure=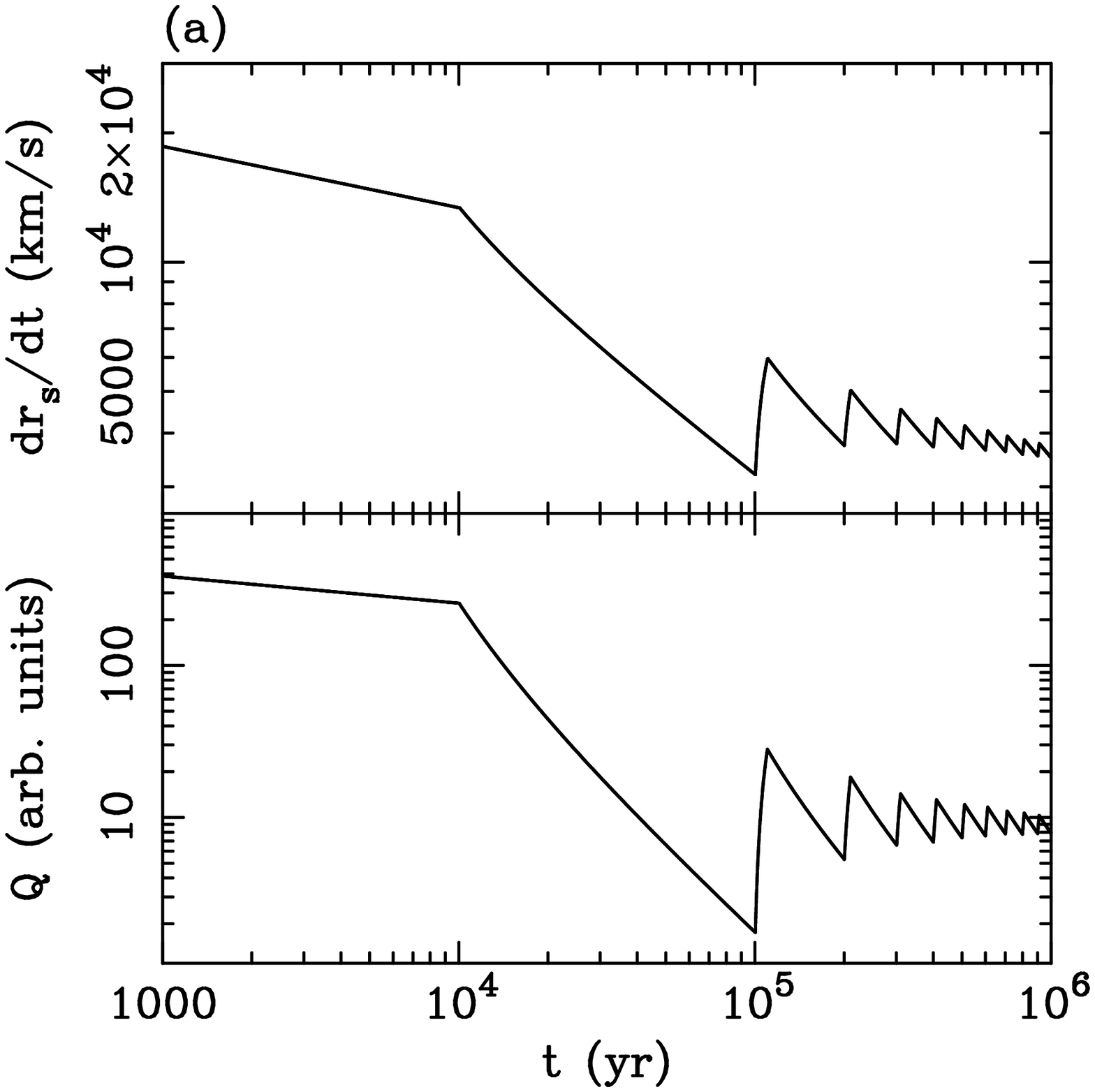,width=0.6\textwidth,angle=270}
\hspace{-2cm}
\psfig{figure=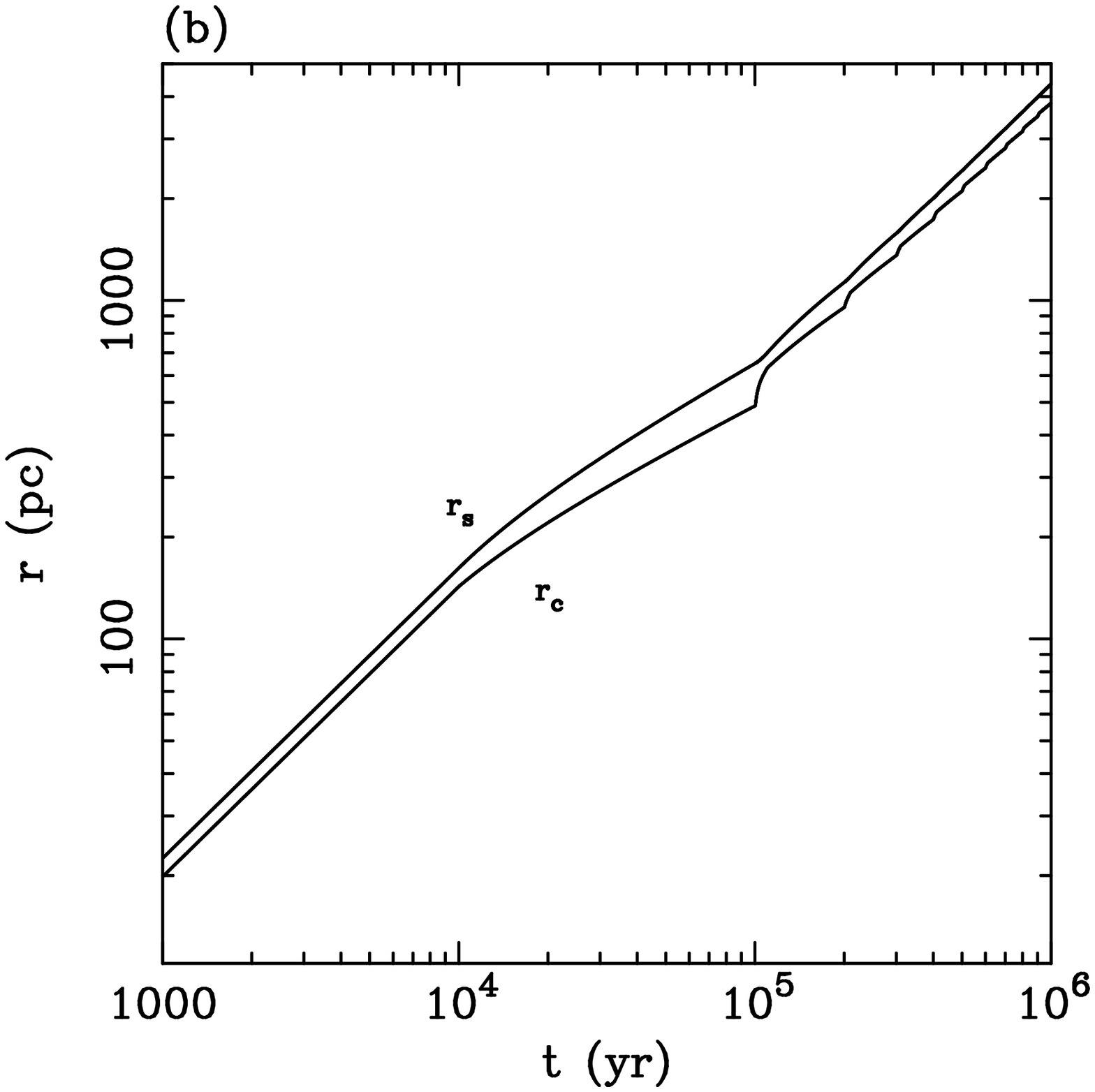,width=0.6\textwidth,angle=270}
}
\caption{Properties of a periodically intermittent source (see Section 2.2
of main text for source parameters).  Panel (a) shows the velocity of the
shock in the ambient medium (top) and the radio luminosity of the cocoon
(bottom).  Panel (b) shows the evolution of the cocoon radius (bottom
curve) and shock radius (top curve).}
\end{figure*}

We now apply the above model to the case of a periodically bursting source.
For concreteness, we shall consider an initially small source which
undergoes 10,000\,yr long bursts that recur every 100,000\,yr.  We take the
jet power to be $L_{\rm j}=10^{46}\ergps$ during the bursts and negligibly
small at other times.  The cocoon is assumed to be dominated by
relativistic material (i.e., $\gc=4/3$) and we take a non-relativistic
equation of state for the shocked ISM/ICM shell (i.e., $\gs=5/3$).
Finally, the following physically reasonable parameters are taken to
characterize the density profile of the ambient medium: $\rho_0=1.7\times
10^{-25}\gpcm$, $a=500\pc$ and $\alpha=1.5$.  The qualitative behaviour
described below is insensitive to reasonable departures from these
canonical parameter values.

Figure~1 shows the results of a numerical integration of eqns. (1)--(2).
During the initial burst of activity ($t<10,000$\,yr) the source expands
self-similarly, i.e., the cocoon radius is a constant multiple of the shock
radius.  This is the evolutionary phase that has been previously examined
by Falle (1991), B96, and Kaiser \& Alexander (1997).  As found in previous
works, $\rs\propto t^{3/(5-\alpha)}$.  By assumption, the power source
switches off at $t=10,000$\,yr and the cocoon/shocked-shell system enters a
`coasting' phase akin to Taylor-Sedov expansion.  It must be stressed that
the expansion during this coasting phase is still pressure-driven, as
opposed to being driven by the momentum of the shocked shell.  Despite the
increased deceleration of the shocked shell, it still remains highly
supersonic for the entire period of this coasting phase (i.e., until
$t=100,000$\,yr).  However, due to the drop in source pressure, the radio
luminosity of the cocoon will fall rapidly once the power source has turned
off.  Thus, even though the basic source structure remains intact, a
coasting source will be significantly fainter than an active source.  It is
also interesting to note that the source evolution is no longer
self-similar in the coasting phase.  

The onset of the second burst of activity induces a rapid increase in
pressure leading to a subsequent increase in both the source expansion rate
and cocoon radio luminosity.  Note that during the (brief) period in which
the contact discontinuity is accelerating (i.e., $\ddot{\rc}>0$), this
interface will be subject to Rayleigh-Taylor instabilities that will tend
to mix material from the shocked shell with the relativistic cocoon
material.  This mixing will produce time-dependence in the effective value
of $\gc$, thereby affecting the detailed evolution of the cocoon.  The
implications of this mixing will be addressed in future work.

After many bursts (or, more precisely, when the recurrence timescale is
short compared with the expansion timescale), the intermittent nature of
the source will be unimportant in determining its evolution.  It will
behave as a constantly fed source with jet power $fL_{\rm j}$, where $f$ is the
fraction of time that the source is on with power $L_{\rm j}$.  The
re-establishment of an approximately self-similar expansion in our
numerical experiment reflects this fact.

\section{Source statistics}

Suppose that we have a single population of evolving radio sources.
Further, suppose we form a flux-limited sample of these sources.  Assuming
a Euclidean universe, the number of sources in the luminosity range
$Q\rightarrow Q+dQ$ is $dN\propto \phi(Q)Q^{3/2}dQ$, where $\phi(Q)dQ$ is
the volume density of sources in that luminosity range.  Since $\phi(Q)dQ$
is proportional to the time that a given source spends in this luminosity
range, we have
\begin{equation}
\phi(Q)\propto \dot{Q}^{-1}=\left(\dot{\rc}\frac{dQ}{d\rc}\right)^{-1},
\end{equation}
implying that
\begin{equation}
\frac{dN}{d\rc}\propto Q^{3/2}\dot{\rc}^{-1}.
\end{equation}
The evolutionary model of Section 2 allows us to determine the functions
$Q(\rc)$ and $\dot{\rc}(\rc)$.  Thus, eqn.~(4) is an explicit expression for
$dN/d\rc$ as a function of $\rc$.  We have chosen to examine the
distribution of $\rc$ since it is the cocoon radius that will be
identified observationally as the half-size of the radio source.

\begin{figure*}[t]
\hbox{
\psfig{figure=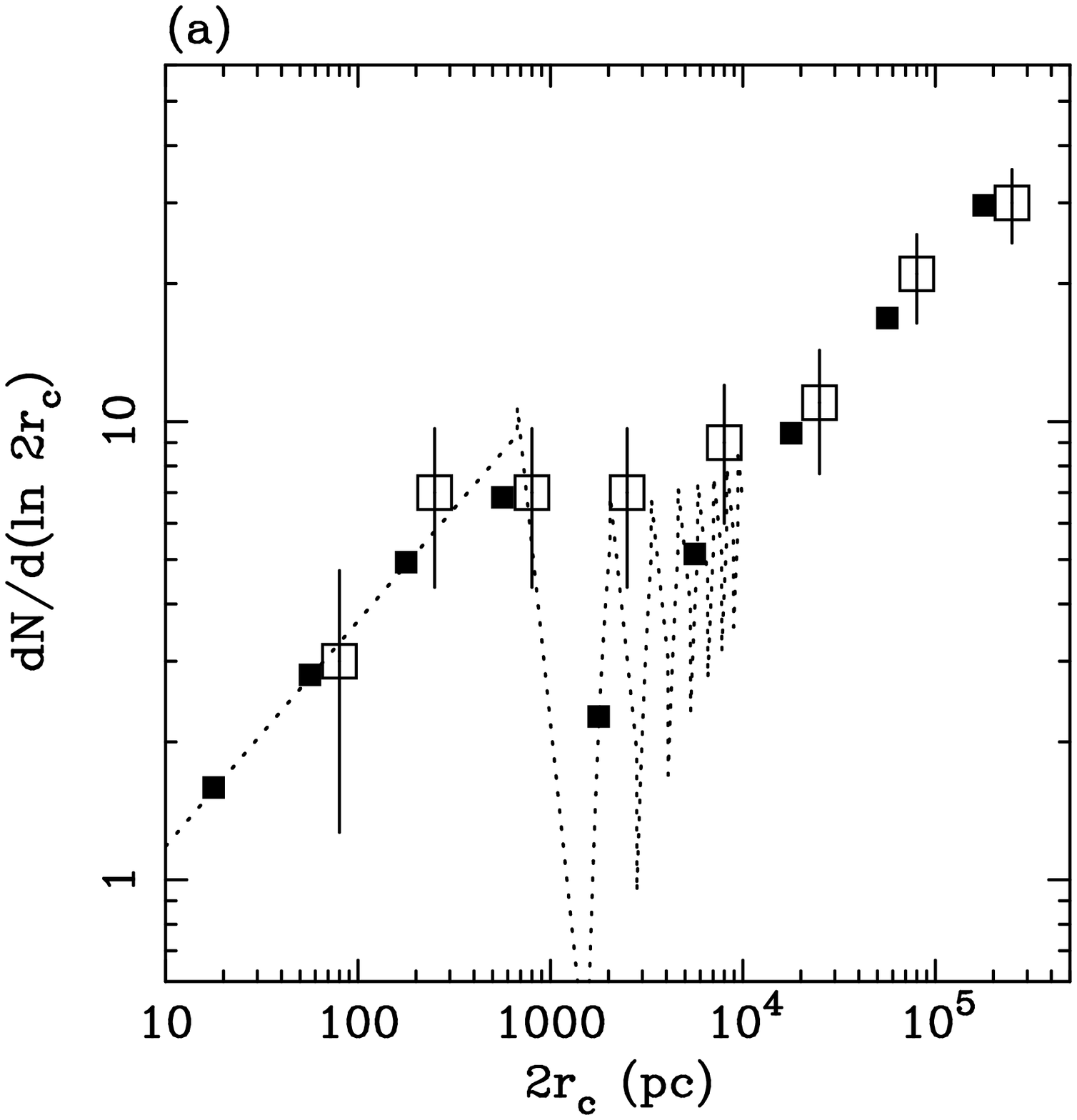,width=0.6\textwidth,angle=270}
\hspace{-2cm}
\psfig{figure=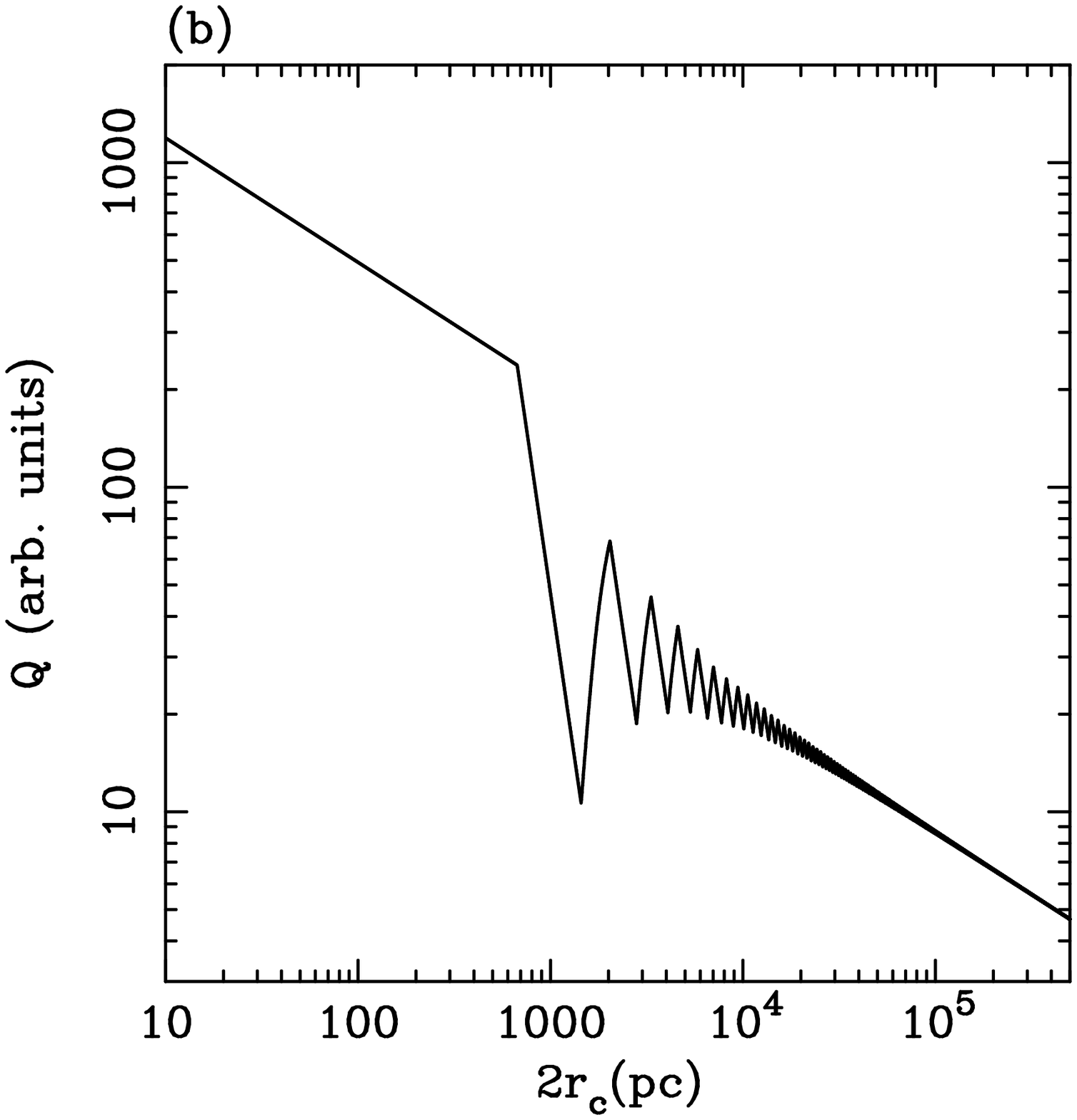,width=0.6\textwidth,angle=270}
}
\caption{(a) Theoretical and observed size distributions.   The dotted line
shows the fine-grained (i.e. unbinned) theoretical model which, for
clarity, has not been displayed beyond $10\kpc$.  The filled squares show
the binned theoretical distribution using bins of size $0.5\log(2\rc)$.
The data from Fig.~10 of OB97 (with 1-$\sigma$ errors) are shown as open
squares.  (b) radio luminosity as a function of $\rc$ for these same
(theoretical) intermittent sources.}
\end{figure*}

Figure 2a shows a comparison of the observed size distribution of OB97 with
our theoretical size distribution for a single population of periodically
intermittent sources.  To facilitate this comparison, we have binned the
theoretical size distribution using bins of $0.5\log(2\rc)$.  In order to
match the distribution of OB97, we set $\alpha=1.8$ (determined by the
slope of the distribution at large sizes) and assume a burst duration of
30,000\,yr.  All other parameters have the values of Section 2.2.

The intermittency of the sources allows the qualitative features of the
OB97 size distribution to be reproduced.  In particular, there is a plateau
in the size distribution resulting from sources that are still undergoing
their first few bursts of activity.  If the break in the OB97 distribution
at small sizes ($\approxlt 100\pc$) is real, this could be identified as
being due to sources that are still undergoing their first burst of
activity.  In this idealized case of a single evolving population, there is
fine structure within the size distribution corresponding to the distinct
cycles of activity.  In practice, the stochastic nature of the parameters
in any real source population will wash out this fine structure.  In
particular, the disagreement between our model and the data at $\sim 2\kpc$
can be resolved if we consider realistic source populations.  Note that we
have not included the largest size bin of the OB97 distribution since this
is probably affected by the complete turning-off of old sources.

Figure 2b shows the radio luminosity $Q$ as a function of the total source
size $2\rc$.  This is to be compared with Fig.~9 of OB97.  An important
feature of Fig.~2b is the dramatic decline in radio luminosity between the
first burst and all subsequent bursts.  In other words, if one were to
assume a constantly fed source and extrapolate from large sources to small
sources, then one would substantially underestimate the small source
luminosity.

\section{Discussion}

There are two immediate predictions of the intermittency model.  First,
Fig.~2b suggests that statistically-complete samples of GPSs and small CSOs
will show them to be very much more radio luminous than suggested by
extrapolation of MSO/FR-II properties.  Assuming that the (optically thin)
radio luminosity depends on cocoon pressure in some way, this prediction
should hold for any physically reasonable prescription relating $p$, $\Vc$
and the radio luminosity, $Q$.  It must be noted, however, that there are
observational complications involved in testing this prediction.  As well
as being comparatively rare, these small sources are often absorbed at
typical radio frequencies.  This may be due to free-free absorption in an
inhomogeneous foreground screen or synchrotron self-absorption in the
source itself.  This absorption must be corrected for before comparisons of
the type discussed above can be made.

A second prediction is that there should be a large number of medium size
objects (a few hundred to a few thousand parsecs across) which are in a
`coasting' phase and have faded below the flux limits of current radio
surveys.  There are several methods that could be employed to search for
such sources.  Deep, low-frequency radio maps might reveal the coasting
cocoons of such sources.  Alternatively, we might hope to observe the
ISM/ICM shock either through X-ray signatures (using the high spatial
resolution of {\it AXAF}) or via the H$\alpha$ emission that it surely
excites (e.g., see Bicknell \& Begelman 1996).

\section{Conclusions}

We have explored the implications of radio source intermittency on source
statistics.  To do this, we have developed a simple model for the evolution
of a cocoon/shocked-shell system which is expanding supersonically into an
ambient medium that possesses a power-law density profile.  The cocoon is
assumed to be fed energy at a rate $L_{\rm j}(t)$.  This model is
integrated numerically for the case of a periodic source which has active
phases (with constant $L_{\rm j}$) separated by inactive, or coasting,
phases in which $L_{\rm j}=0$.  During the first few periods of inactivity,
the radio luminosity will fade rapidly.  However, these young sources can
maintain highly supersonic expansion during their coasting phases and,
hence, will remain intact throughout the inactive periods.  Once a source
has grown large enough such that the expansion timescale is longer than the
recurrence timescale, the intermittency will not affect its subsequent
evolution.  The fading of small, inactive sources, and the effect of
intermittency on the expansion velocity of the sources, produce a
double-break in the size distribution.  This can be readily identified with
the plateau found in the size distribution of OB97.

There are two clear predictions of this model.  First, one could search for
the faint, inactive sources.  These sources might reveal themselves in
deep, low-frequency radio surveys.  Alternatively, they could be detectable
via the X-ray signatures or H$\alpha$ emission accompanying the coasting
ISM/ICM shock front.  Secondly, statistically complete radio samples of
small sources ($\sim 100\pc$ and smaller), once corrected for absorption
effects, should show these sources to be very overluminous as compared with
an extrapolation from larger sources.  Since this is essentially reflecting
the high pressure of these small sources, this prediction should be
independent of the precise form of the radio emissivity.

\section*{Acknowledgments}

This work has been supported by the National Science Foundation under grant
AST-9529175.

%\end{multicols}


\begin{thebibliography}{}
\bibitem[]{} Begelman M.~C., 1996, in Proceedings of Cygnus-A: Study of a
Radio Galaxy, eds C.Carilli \& D.Harris (Cambridge University Press,
Cambridge), p.209 (B96)
\bibitem[]{} Begelman M.~C., Cioffi D.~F., 1989, ApJ, 341, 685 
\bibitem[]{} Bicknell G.~V., Begelman M.~C., 1996, ApJ, 467, 597
\bibitem[]{} Carilli C.~L., Perley R.~A., Harris D.~E., 1994, MNRAS, 270,
173
\bibitem[]{} Falle S.~A.~E.~G., 1991, MNRAS, 250, 581 
\bibitem[]{} Fanti C., Fanti R., Dallacasa D., Schilizzi R.~T., Spencer R.~E.,
Stanghellini C., 1995, A\&A, 302, 317
\bibitem[]{} Kaiser C.~R., Alexander P., 1997, MNRAS, 286, 215
\bibitem[]{} O`Dea C.~P., Baum S.~A., 1997, AJ, 113, 148 (OB97)
\bibitem[]{} O'Dea C.~P., Baum S.~A., Stanghellini C., 1991, ApJ, 380, 66
\bibitem[]{} Readhead A.~C.~S., Taylor G.~B., Pearson T.~J., Wilkinson P.~N.,
1996, ApJ, 460, 634
\bibitem[]{} Scheuer P.~A.~G., 1974, MNRAS, 166, 513
\bibitem[]{} Wilkinson P.~N., Polatidis A.~G., Readhead A.~C.~S., Xu W.,
Pearson J.~J., 1994, ApJ, 432, L87
\end{thebibliography}
\end{document}